\documentstyle[aps,prl,psfig]{revtex}

\newcommand{\be}{\begin{equation}}
\newcommand{\ee}{\end{equation}}
\newcommand{\bea}{\begin{eqnarray}}
\newcommand{\eea}{\end{eqnarray}}

\tightenlines

\begin{document}

\draft
\twocolumn[\hsize\textwidth\columnwidth\hsize\csname 
@twocolumnfalse\endcsname

\title{\bf Phase diagram for a class of spin-half Heisenberg models\\
interpolating between the square-lattice, the triangular-lattice\\
and the linear chain limits}
\author{Zheng Weihong\cite{byline1} and Ross H. McKenzie\cite{byline2}} 
\address{School of Physics,                                              
The University of New South Wales,                                   
Sydney, NSW 2052, Australia.}                      
\author{Rajiv R. P. Singh\cite{byline3}}
\address{Department of Physics,
University of California,
Davis, CA 95616}

%\date{Dec. 5, 1998}
\date{\today}

\maketitle 

\begin{abstract}
We study the spin-half Heisenberg models on an anisotropic 
two-dimensional lattice
which interpolates between the square-lattice at one end, a set
of decoupled spin-chains on the other end, and the triangular-lattice
Heisenberg model in between. By series expansions around two different
dimer ground states
and around various commensurate and incommensurate
magnetically ordered states, we establish the phase diagram for this model
of a frustrated antiferromagnet.
We find a particularly rich phase diagram due to the interplay
of magnetic frustration, quantum fluctuations and varying dimensionality.
%as a function of the amount of magnetic frustration.
There is a large region of the usual 2-sublattice Ne\'el phase, a 3-sublattice
phase for the triangular-lattice model, a region of incommensurate
magnetic order around the triangular-lattice model, and regions in parameter
space where there is no magnetic order. We find that the incommensurate ordering
wavevector is in general altered from its classical value by
quantum fluctuations. The regime of weakly coupled chains is particularly
interesting and appears to be nearly critical.\\
\\
\end{abstract}                                                              

\phantom{.}
]

\narrowtext
\section{INTRODUCTION}

Quantum phases and phase transitions in 
two-dimensional Heisenberg models
are of great current interest.\cite{chn-csy}
Following much experimental, theoretical, and numerical work, a 
fairly comprehensive and consistent picture
has emerged for the ordered phases
of unfrustrated Heisenberg models at zero and finite temperatures
as well as for the quantum
critical points separating them from the quantum
 disordered phases.\cite{auerbach,sachdev0} The case of frustrated Heisenberg
models, potentially includes much richer phenomena and is relatively
less understood. Here, the lattice models that have been studied the most
are the square-lattice nearest and second neighbor interaction
(or $J_{nn}-J_{nnn}$) model, where as a function of $J_{nnn}/J_{nn}$, the
two and four sublattice colinear Ne\'el phases are separated by
a region of quantum disordered ground states.\cite{zheng}
Theoretical arguments and various numerical results suggest that
this intermediate phase is spontaneously dimerized. Other models
that have received considerable attention are the triangular-lattice and
Kagome-lattice Heisenberg models.\cite{rajiv_triangle,kagome-fr,kagome-no}
 The former appears to be
magnetically ordered in the three-sublattice non-colinear pattern,
whereas the latter almost surely has a non-magnetic ground state.
The nature of the magnetically disordered phases and low energy
excitations in these frustrated Heisenberg models are subjects
of considerable interest.

Here we study a class of Heisenberg models that interpolate between
the square-lattice, the triangular-lattice and the linear chain Heisenberg
models. It can be defined in terms of a square-lattice of spin-half
operators, with a nearest neighbor Heisenberg interaction $J_2$ and a 
second neighbor interaction $J_1$ along only one of the diagonals of
the lattice. In the limit $J_1\to 0$, we recover the square-lattice
Heisenberg model. In the limit $J_2\to 0$, the model reduces to
decoupled spin-chains running along the diagonals of the square-lattice.
For small $J_2/J_1$ the model consists of weakly coupled chains with
frustrated interchain coupling.
For $J_2=J_1$, the model is exactly equivalent to the triangular-lattice
Heisenberg model. 

This model is of particular interest for at least two reasons.
First, within a single Hamiltonian, it provides a way of interpolating
between several well-known one and two dimensional models. One can
study the role of frustration in going from one to two dimensions,\cite{ssf}
as well as the stability of commensurate and incommensurate
magnetic order with respect to quantum fluctuations.\cite{chandra}
Second, the model is of direct relevance to the magnetic
phases of various
quasi-two-dimensional organic superconductors.
It has been argued that this model should describe
the spin degrees of freedom of the insulating
phase of the layered molecular crystals,
$\kappa$-(BEDT-TTF)$_2$X.\cite{4mck}
These materials probably have $ J_1/J_2 \sim 0.3 - 1$.
This ratio varies with the anion X and should vary with uniaxial
stress applied within the layer along the diagonal.\cite{campos}
A very slight variant on this model with $ J_1/J_2 \sim 4 $
has been proposed to describe 
the insulating phase of $\theta$-(BEDT-TTF)$_2$RbZn(SCN)$_4$
which has a spin gap.
Recent studies of the associated Hubbard model on an
anisotropic triangular lattice at half-filling
suggest that the wave vector associated with the
spin excitations determines the type of superconducting
order.\cite{flex}
Consequently, it appears that a good understanding of
this Heisenberg model may be a pre-requisite to understanding
organic superconductivity.

In this paper we study the zero-temperature
 phase diagram of this model by series
expansions around various dimerized and magnetically ordered phases.
Because of frustration, and large coordination number of the lattice,
it would be difficult to get a comparable treatment of this model
by any other numerical method.
We will only consider $J_1>0, J_2>0$, that is both couplings are
antiferromagnetic, which introduces frustration into the model.
We find that the Ne\'el order of the square-lattice Heisenberg
model persists up
to $J_1/J_2\lesssim 0.7$, where it goes continuously
to zero. It is interesting to note that Ne\'el order exists
beyond it regime of classical stability, providing an example
of quantum order by disorder. \cite{chandra}
In the region $0.7\lesssim J_1/J_2 \lesssim 0.9$,
there is no magnetic order and the ground-state
is dimerized in the columnar pattern. For larger values of $J_1/J_2$, there is
incommensurate or spiral long-range order. Except for the
triangular-lattice point ($J_1/J_2=1$), we find that the 
ordering wavevector is different from the classical Heisenberg
model. At the triangular-lattice point, the dimer expansions 
indicate a magnetic instability at 
an ordering wavevector which coincides with the classical value.
Furthermore, our Ising expansion series in this case
reduce to previously derived series expansions directly
for the triangular-lattice model.\cite{rajiv_triangle}

The region of large $J_1/J_2$ is very interesting and is not fully
resolved by our series expansions. In the limit of small $J_2$, we
have Heisenberg spin-chains coupled by frustrating zig-zag
interactions. For just two such chains, this problem was
studied by Affleck and White.\cite{bursill}
 This zig-zag ladder has incommensurate
spin correlations, a small gap and a spontaneous dimerization,
which breaks the degeneracy of the weak zig-zag bonds but leaves
the strong chain bonds equivalent. Our numerical results suggest
that such a state is definitely not favored for the 
two-dimensional system.
This is not hard to understand as such a broken symmetry phase
can only have the strong bonds between every alternate pair of chains.
We suggest three possible scenarios for this phase: (i) It
has weak incommensurate spiral order, which vanishes rapidly
(with a power greater than unity)
as $J_2\to 0$, (ii) This phase is dimerized along the strong
$J_1$ bonds and has a spin-gap, (iii) The phase remains critical
without any long-range order. The ground state energy obtained
from the Ising expansions around a spiral phase and
around a dimer state along the strong bonds remain nearly equal
through much of this phase. In all three
cases the system is close to being critical. 

The elementary excitations
of this phase and how they are connected to the excitations in
isolated spin-chains should be very interesting to investigate.\cite{spectral}
The low-energy properties of models
with non-collinear order are
described by the SO(3) non-linear sigma model.\cite{chubukov,angel}
In the quantum disordered phase of this model
the elementary excitations are deconfined spin-$1/2$ objects,
%This is consistent with the predictions of large
%$N$ treatments of Sp($N$) non-collinear 
%antiferromagnets.\cite{read,sachdev0}
or spinons,\cite{read,sachdev0}
in contrast to the magnons or spin-one triplets that are
the elementary excitations in most two-dimensional 
antiferromagnets. Thus, this model, with spiral classical order
provides us with candidate
systems to look for spinon excitations.

The plan of the paper is as follows. In Sec. II %the next few sections, 
we discuss the model and the various series expansions that are
carried out. % Following that 
In Sec. III we discuss
the results for the phase diagram, the ground state energy,
the excitation spectra and the correlation functions of the model. Finally,
in the last section we present our conclusions and discuss some future
directions.

%=======================================================================
\begin{figure}[h] %h: here; t:top of page; b:bottom of page; p: page of float
%\vspace{9pt}
\par
\centerline{\hbox{\psfig{figure=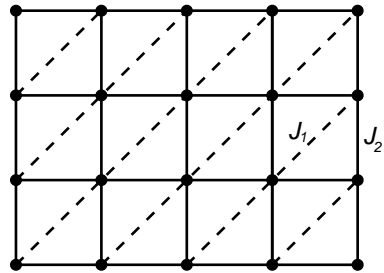,width=5cm}}}
%\par
\vspace{10pt}
\caption{The anisotropic triangular lattice with coupling $J_1$ and $J_2$.
}
\label{fig_lat}
\end{figure}
%=======================================================================

\section{Series Expansions}

The Heisenberg antiferromagnet on an anisotropic  triangular lattice is
equivalent to the Heisenberg antiferromagnet on a square lattice with 
nearest neighbor interactions and one of diagonal next-nearest neighbor
interactions as shown in Fig. 1. The 
Hamiltonian is:
\begin{equation}
H = J_1 \sum_{\langle in\rangle } {\bf S}_i\cdot {\bf S}_n 
+ J_2 \sum_{\langle ij\rangle} {\bf S}_i\cdot {\bf S}_j \label{H}
\end{equation}
where $\langle ij\rangle$ means pairs of nearest neighbors and
$\langle in\rangle$ means one of diagonal next-nearest neighbor pairs, as
illustrated in Fig. 1. 
We denote the ratio of the couplings as $y$, that is, $y \equiv J_1/J_2$.
In the present paper, we study only the case of antiferromagnetic coupling, 
where both $J_1$ and $J_2$ are positive.
In the limit $J_2=0$, the model is  equivalent
to decoupled one-dimensional Heisenberg chains.
In the limit $J_1=0$, the model is  equivalent
to the two-dimensional square-lattice Heisenberg model.
For $J_1=J_2$, the model is equivalent to the Heisenberg model
on a triangular lattice.

\begin{figure}[h] %h: here; t:top of page; b:bottom of page; p: page of float
\vspace{-1cm}
%\par
\centerline{\hbox{\psfig{figure=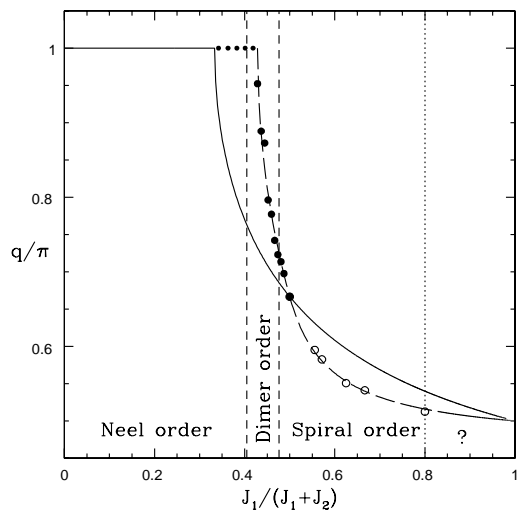,width=9cm}}}
\par
%\vspace{-39pt}
\caption{Phase diagram of the model. The wavevector
$q$ is plotted as a function of $J_1/(J_1+J_2)$. The solid curve is the classical
ordering wavevector, while the points are estimates of the location
of the minimum triplet gap from both columnar dimer expansions (solid points) 
and diagonal dimer expansions (open points).
The long dashed lines are the fit given in Eq.  (\protect\ref{eq_qfit}). 
The vertical short dashed lines indicate the region of spontaneously dimerized
phase, and vertical dotted line indicate another possible transition point.
}
\label{fig_q}
\end{figure}

Let's discuss the classical ground state of this system:
If we assume $\langle {\bf S}_i\rangle$ lies in the $x-z$ plane, 
we can rotate all the spins related to a reference spin so we
have a ferromagnetic ground state.
The classical ground state of this system has nearest neighbor spins differ
by an angle $q$, and spin joined by a diagonal bond differ by $2q$, 
where $q$ is
\begin{equation}
q = \cases {\pi, & $J_1\le J_2/2$;\cr
\arccos (-J_2/2J_1), &$J_1 > J_2/2$.\cr} \label{eq_q}
\end{equation}
The angle $q$ as a function of $J_1/(J_1+J_2)$ is shown as
the solid curve in Fig. \ref{fig_q}.

After this rotation, the transformed Hamiltonian is
\begin{equation}
H = H_1 + J_1 H_2 + J_2 H_3 ,  \label{Hq}
\end{equation}
where
\begin{eqnarray}
H_1 &=&  J_1 \cos (2q) \sum_{\langle in\rangle} S_{i}^z S_{n}^z
+ J_2 \cos (q) \sum_{\langle ij\rangle } S_{i}^z S_{j}^z  ~, \nonumber \\
H_2 &=& \sum_{\langle in\rangle} S_{i}^y S_{n}^y + \cos(2q) S_{i}^x S_{n}^x 
   + \sin (2q) ( S_{i}^z S_{n}^x - S_{i}^x S_{n}^z ) \label{Hq012} \\
H_3 &=& \sum_{\langle ij\rangle} S_{i}^y S_{j}^y + \cos(q) S_{i}^x S_{j}^x 
   + \sin (q) ( S_{i}^z S_{j}^x - S_{i}^x S_{j}^z )~. \nonumber 
\end{eqnarray}

We have studied this system by using 
linked-cluster expansion methods at zero-temperature 
including Ising expansions, and dimer expansions about 
two different dimerization patterns.
% the vertical and horizontal bonds.
The linked-cluster expansion method
has been previously reviewed in several
articles\cite{he90,gel90,gelmk}, 
and will not be repeated here.
Here we will only summarize
the expansion methods used, and the results derived from them 
will be presented.
The series coefficients are not given here,
but are available upon request.
Previous studies using these methods have agreed quantitatively
with other numerical studies (including quantum Monte Carlo and exact
diagonalization) for unfrustrated spin models. For frustrated
two-dimensional models, both the Monte Carlo and the exact diagonalization
methods face problems (due to the minus signs in case of the
former and tremendous variations with size and shapes of
the clusters in case of the latter). Thus, we believe, it would
be difficult to get a comparable numerical treatment of this strongly
frustrated model by other numerical methods.

\subsection{Ising Expansions}
To construct a $T=0$ expansion about 
the classical ground state for this system,
one has to introduce an anisotropy parameter $\lambda$, 
and write the Hamiltonian for the Heisenberg-Ising model as
\begin{equation}
H = H_0 + x V~,  \label{Hising}
\end{equation}
where
\begin{eqnarray}
H_0 &=&  H_1 - t \sum_{i }  S_{i}^z  ~, \nonumber \\ 
 && \\
V &=& J_1 H_2 + J_2 H_3 + t \sum_{i}  S_{i}^z   \nonumber 
\end{eqnarray}
The last term of strength $t$
in both $H_0$ and $V$ is a local field term,
% (staggered in the original variables), 
which can be included to improve convergence. 
The limits $x =0$ and $x =1$ correspond to the Ising model and
the isotropic Heisenberg model, respectively.
The operator $H_0$ is taken as the unperturbed
Hamiltonian, with the unperturbed ground state being the
usual ferromagnetically ordered state.
The operator $V$ is treated as a perturbation.
%It flips a pair of spins on neighboring sites.

%=======================================================================
%=======================================================================

To determine whether the system is in the ordered phase, we need to
compute the order parameter $M$ defined by
\begin{equation}
M = {1\over N} \sum_i \langle S_i^z\rangle
\end{equation}
where the brackets denote ground-state expectation values.
Actually, in this expansion, we can choose $q$ different from
that given in Eq. (\ref{eq_q}), the favored $q$ should give a lower
energy. If $q=\pi$, we are expanding about the usual Ne\'el phase.

Ising series have been calculated for
the ground state energy per site, $E_0/N$,
the order parameter $M$, 
and the triplet excitation spectrum (for the case of $q=\pi$ only) 
%or the energy gap at ${\bf k}=0$
for several ratios of couplings $x$ and (simultaneously) for several 
values of $t$ up to order $x^{10}$.
This calculation involves 116137 linked-clusters of up to 10 sites. 
The series for the case of $y=1$ have been computed
previously\cite{rajiv_triangle},
and our results agree with these previous results.
This is a highly non-trivial check on the correctness of
the expansion coefficients, as the set of 
(anisotropic square-lattice) graphs considered
here are completely different from the 
(triangular-lattice) graphs considered in
Ref. \cite{rajiv_triangle} and the representation of
the Hamiltonian as a Heisenberg-Ising model is also
superficially different.

At the next stage of the analysis, we try to extrapolate
 the series to the isotropic point ($x=1$)
for those values of the exchange coupling
parameters which lie within the ordered phase at $x=1$.
For this purpose, we first transform the series to a new variable
\begin{equation}
\delta = 1- (1-x)^{1/2}~,
\end{equation}
to remove the singularity at $x=1$ predicted by the spin-wave theory. This
was first proposed by Huse\cite{hus} and was also
used in earlier work on the square lattice case\cite{zhe1}.
We then use both integrated first-order inhomogeneous
differential approximants\cite{gut} and  Pad\'{e} approximants to
extrapolate the series to the isotropic point $\delta=1$  ($x=1$).
The results of the Ising expansions will be presented in later sections.

%=======================================================================
\begin{figure}[h] %h: here; t:top of page; b:bottom of page; p: page of float
%\vspace{9pt}
\par
\centerline{\hbox{\psfig{figure=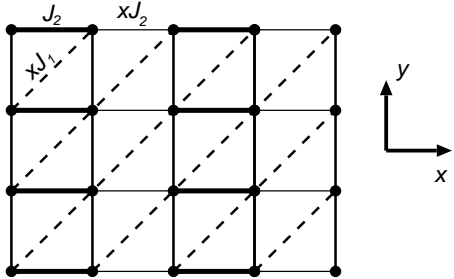,width=6cm}}}
%\par
\vspace{10pt}
\caption{Columnar dimerization pattern, the thick solid lines represent strongest
bonds $J_2$, and thin solid lines and dashed lines 
represent weaker bonds $xJ_2$ and $xJ_1$, respectively. 
Note that the dimer centers lie
on a rectangle lattice with spacing $2a$ in $x$-direction and $a$ 
in $y$-direction.
In characterizing the excitation spectrum
we set the lattice spacing in both the $x$- and $y$-direction to be 1.
}
\label{fig_lattice_cdimer}
\end{figure}
%=======================================================================

\subsection{Dimer Expansions}
\begin{figure}[h] %h: here; t:top of page; b:bottom of page; p: page of float
%\vspace{9pt}
\par
\centerline{\hbox{\psfig{figure=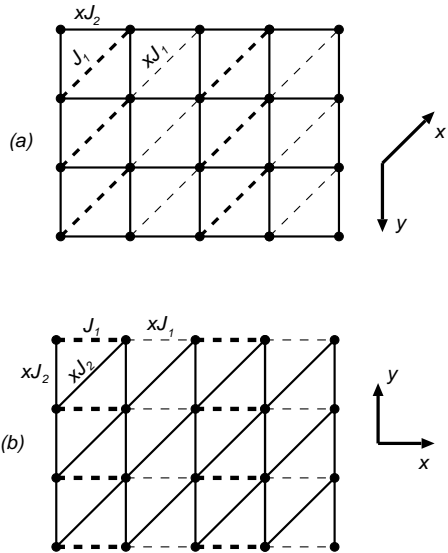,width=6cm}}}
%\par
\vspace{10pt}
\caption{(a) Diagonal dimerization pattern, the thick dashed lines 
represent strongest bonds $J_1$, and thin dashed lines and solid lines 
represent weaker bonds $xJ_1$ and $xJ_2$, respectively. 
(b) The equivalent lattice to (a) used in series expansions.
Again note that the dimer centers lie
on a rectangle lattice with spacing $2a$ in $x$-direction and $a$ in $y$-direction.
In characterizing the excitation spectrum
we set the lattice spacing in both $x$- and $y$-direction to be 1.
}
\label{fig_lattice_ddimer}
\end{figure}

For this system, we have carried out two types of  dimer expansions 
corresponding to two types of dimer coverings. 
One is the columnar covering displayed in Fig. \ref{fig_lattice_cdimer},
and the other is the diagonal covering 
displayed in Fig. \ref{fig_lattice_ddimer}(a).

\subsubsection{Columnar dimer expansions}
In this expansion, we take the columnar bonds (D) denoted by the 
thick solid bonds in Fig. \ref{fig_lattice_cdimer} as the unperturbed Hamiltonian $H_0$, 
and the rest of the bonds as perturbation, that is, we define
the  following ``dimerized $J_1-J_2$ model"
\begin{equation}
H = J_2 \sum_{(i,j)\in D} {\bf S}_i \cdot {\bf S}_j + 
x {\Big [} J_2 \sum_{(i,j)\not\in D} {\bf S}_i \cdot {\bf S}_j +
J_1 \sum_{\langle in\rangle} {\bf S}_i \cdot {\bf S}_j {\Big ]}
\end{equation}
We take the first term in above $H$ as the unperturbed Hamiltonian, and
the second term as perturbation. 
The unperturbed ground state is a product state of nearest neighbor
singlet dimers and the perturbation couples these among themselves and
with the pair-triplet states. At $x=1$, one recovers the 
original Hamiltonian  (\ref{H}).

Two types of series can be obtained in this way. By fixing  $y$,
we can compute series in the single variable $x$ up to order $L$,
\begin{equation}
E/J_2 = \sum_{i=0}^L c_i(y) x^i
\end{equation}
In the other approach we keep both $x$ and $y$ as 
expansion parameters and obtain double series of the form
\begin{equation}
E/J_2 = \sum_{i=0}^L \sum_{j=0}^i d_{ij} (y)^i x^j
\end{equation}
where the coefficients $d_{ij}$ are computed up to order
$i=L$. 

The series has been computed
up to order $L=9$ for
the ground-state energy $E_0$, and up to order $L=8$ 
for the triplet excitation spectrum $\Delta (k_x,k_y)$.
This calculation involves 49684 linked clusters of up to 9 sites. 

Note that if the centers of the dimers are treated as lattice sites,
one obtains a rectangular lattice with spacing $2a$ in $x$-direction 
and $a$ in $y$-direction, 
where $a$ is the lattice spacing of original lattice.
In characterizing the excitation spectrum presented in the later sections
we set the lattice spacing in both x- and y-directions to be 1.
In this notation, 
the wavevector corresponding to the classical ordering is
$(q-\pi,q)$, which varies from $(0,\pi)$ to $(-\pi/2,\pi/2)$ as
$J_1/J_2$ increases from 0 to $\infty$.

The leading order series for the triplet excitation spectrum is:
\begin{eqnarray}
\Delta (k_x,k_y)/J_2 && = 1 + x [ - \case 1/2 \cos(k_x) + (1- J_1/2J_2) \cos(k_y) \nonumber \\
&& -  (J_1/2J_2) \cos(k_x+k_y)] + O(x^2)
\end{eqnarray}
Note that the minimum of the above spectrum may not coincide with
the ordering wavevector $(q-\pi,q)$ of the classical system.

%=======================================================================
%=======================================================================

\subsubsection{Diagonal dimer expansions}
In this expansion, we take the diagonal bonds (D) denoted by the 
thick dashed bonds in Fig. \ref{fig_lattice_ddimer}(a) as 
unperturbed Hamiltonian $H_0$, and rest of the bonds as 
perturbation. That is we define the following ``dimerized $J_1-J_2$ model"
\begin{equation}
H = J_1 \sum_{(i,n)\in D} {\bf S}_i \cdot {\bf S}_j + 
x {\Big [} J_1 \sum_{(i,n)\not\in D} {\bf S}_i \cdot {\bf S}_j +
J_2 \sum_{\langle ij\rangle} {\bf S}_i \cdot {\bf S}_j {\Big ]}
\end{equation}
Again at $x=1$, one recovers the 
original Hamiltonian (\ref{H}).

For convenience, we have transformed the lattice shown in Fig. \ref{fig_lattice_ddimer}(a)
to the equivalent one shown in Fig. \ref{fig_lattice_ddimer}(b).
Again the dimer centers in Fig. \ref{fig_lattice_ddimer}(b)  lie
on a rectangular lattice with spacing $2a$ in $x$-direction and $a$ in $y$-direction. In
characterizing the excitation spectrum presented in the later sections
we also set the lattice spacing in both $x$- and $y$-directions to be 1.
In this representation 
the wavevector corresponding to the classical ordering  is
$(\pi-2q ,q)$, which varies from $(-\pi,\pi)$ to $(0,\pi/2)$ as
$J_1/J_2$ increases from 0 to $\infty$.

As in the case of columnar dimer expansions, 
two types of series have been computed 
up to order $x^9$ for
the ground-state energy $E_0$, and up to order $x^8$ 
for the triplet excitation spectrum $\Delta (k_x,k_y)$.
This calculation involves the same linked clusters as the 
columnar dimer expansions.

The leading order series for triplet excitation spectrum is
\begin{eqnarray}
\Delta(k_x,k_y) && = 1 + x [ - {\case 1/2} \cos(k_x) + (J_2/2J_1) \cos(k_y) \nonumber \\
&& - (J_2/2J_1) \cos(k_x+k_y) ] + O(x^2)
\end{eqnarray}
Note that the minimum of $\Delta$ is at $(\pi - 2 q, q)$, with $q$ defined
in Eq. (\ref{eq_q}), that is just the ordering wavevector of the classical system.

\section{Results}
Having obtained the series for the various expansions above we present in this section
the results of series analysis. We use 
integrated first-order inhomogeneous
differential approximants\cite{gut} and  Pad\'{e} approximants
to extrapolate the series. 

%=======================================================================
\begin{figure}[h] %h: here; t:top of page; b:bottom of page; p: page of float
\vspace{-1cm}
%\par
\centerline{\hbox{\psfig{figure=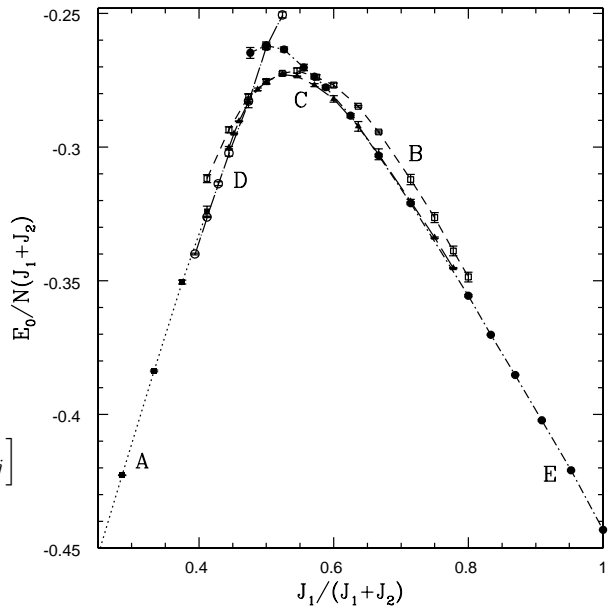,width=10cm}}}
\par
\vspace{-1.3cm}
\caption{The ground-state energy per site $E_0/N(J_1+J_2)$  as
function of $J_1/(J_1+J_2)$, obtained from the Ising expansions about
the Ne\'el order (i.e. $q=\pi$) (A), the Ising expansions about the 
classical spiral order (i.e. $q\not= \pi$ defined by Eq. (\protect\ref{eq_q}) (B),
the Ising expansions using the $q$ obtained by dimer expansions (C),
the columnar dimer expansions (D) and the diagonal dimer expansion (E).
The result at $J_2=0$ is the exact results for the ground state energy
of a antiferromagnetic spin-chain\protect\cite{chainexact}.
}
\label{fig_e0}
\end{figure}
%=======================================================================

\subsection{Ground state energy}

Fig. \ref{fig_e0} shows the ground state energy per site $E_0/N$ 
for the Hamiltonian in Eq. (\ref{H})
obtained from various expansions,
We see that the Ising expansions about the Ne\'el order 
give the lowest energy and also show the best convergence in the 
region of $J_1/J_2 \lesssim 0.7$.
The Ising expansions about the classical ground state
(i.e. $q\not= \pi$ defined by Eq. (\ref{eq_q})) gives the lowest energy
in the region of $0.9 \lesssim J_1/J_2 \lesssim 1.2$, 
the columnar dimer expansions gives the lowest energy for
 $0.65 \lesssim J_1/J_2 \lesssim 0.9$, and the diagonal
dimer expansions
give the lowest energy for $J_1/J_2 \gtrsim 1.2$. Whenever the
dimer expansions indicate an ordering wavevector different from
the classical one, we have also computed the series expansion
around the spiral-state with this modified wavevector. These
always give lower ground state energies than those for the
classical wavevector. Also, for $J_1/J_2\gtrsim 1.2$, these
modified Ising expansions lead to ground state energies
that are almost identical with the diagonal dimer expansions.

%=======================================================================
\begin{figure}[h] %h: here; t:top of page; b:bottom of page; p: page of float
\vspace{-1cm}
%\par
\centerline{\hbox{\psfig{figure=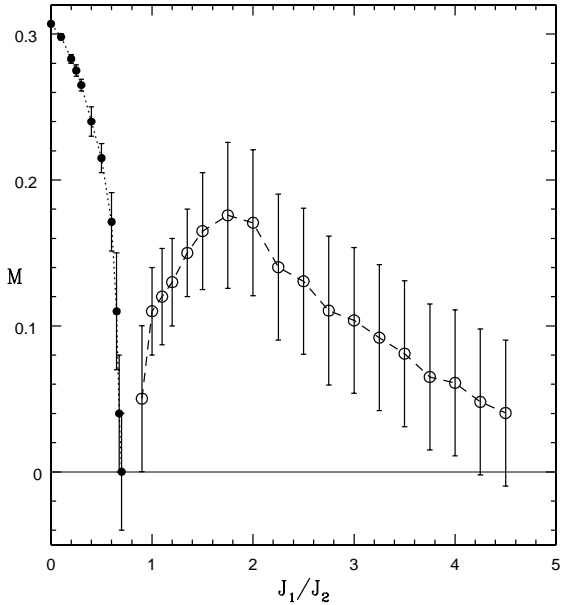,width=10cm}}}
%\par
\vspace{-1.3cm}
\caption{The order parameter $M$ 
{\it versus} $J_1/J_2$.
The solid points with error bars are the estimates from
the Ising expansions about Ne\'el order (i.e. $q=\pi$) , while the open
points are the estimates from Ising expansion about the 
classical spiral order (i.e., $q$ defined by Eq. (\protect\ref{eq_q})).
}
\label{fig_M}
\end{figure}
%=======================================================================

\subsection{Magnetisation}

The order parameter $M$ obtained from the Ising expansions
about the classical ground state is shown in
Fig. \ref{fig_M}. 
The dependence on the amount of magnetic frustration
is very similar to that obtained from linear spin
wave theory.\cite{merino}
The staggered magnetization associated with 
 the Ne\'el order
exists up to $J_1/J_2 \simeq 0.7$, rather than up to $J_1/J_2=0.5$, as
in the classical system.
 We can also see that the magnetisation $M$ vanishes over
the region $0.7 \lesssim J_1/J_2 \lesssim 0.9$, suggesting that this
may belong to a quantum disordered phase. For $J_1/J_2 \gtrsim 0.9$,
the system may have spiral order.

The fact that the Ne\'el order survives beyond the
classicaly stable region, is an example of promotion of
collinear order by quantum fluctuations. A similar stabilization
was found in the 1/4 depleted square-lattice Heisenberg model
relevant to the material CaV$_3$O$_7$ by Kontani et al.\cite{kontani} This
can be viewed as a part of the more general phenomena of quantum order by
disorder,\cite{chandra} where quantum fluctuations help select and
stabilize an appropriate type of order, typically collinear, in face of near
classical degeneracy.

%=======================================================================
\begin{figure}[t] %h: here; t:top of page; b:bottom of page; p: page of float
\vspace{-1.5cm}
%\par
\centerline{\hbox{\psfig{figure=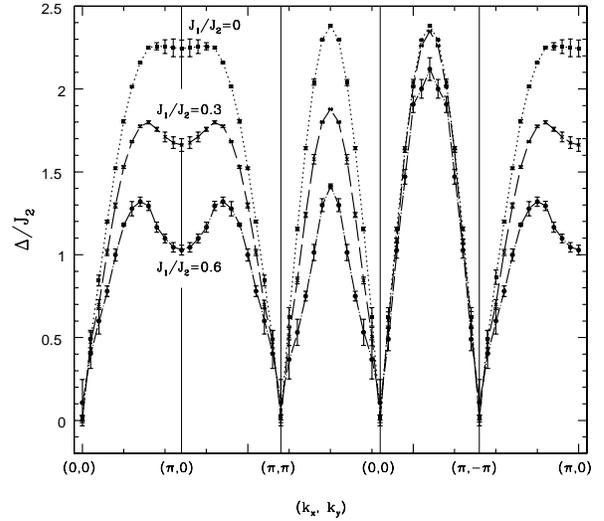,width=9cm}}}
%\par
\vspace{-1.5cm}
\caption{Plot of 
triplet excitation  spectra
$\Delta(k_x, k_y)$ 
(derived from the Ising expansions about Ne\'el order) along high-symmetry
cuts through the Brillouin zone for  coupling ratios
$J_1/J_2=0, 0.3, 0.6$.
}
\label{fig_Ising_mk}
\end{figure}
%=======================================================================

%=======================================================================
\begin{figure}[b] %h: here; t:top of page; b:bottom of page; p: page of float
\vspace{-1cm}
%\par
\centerline{\hbox{\psfig{figure=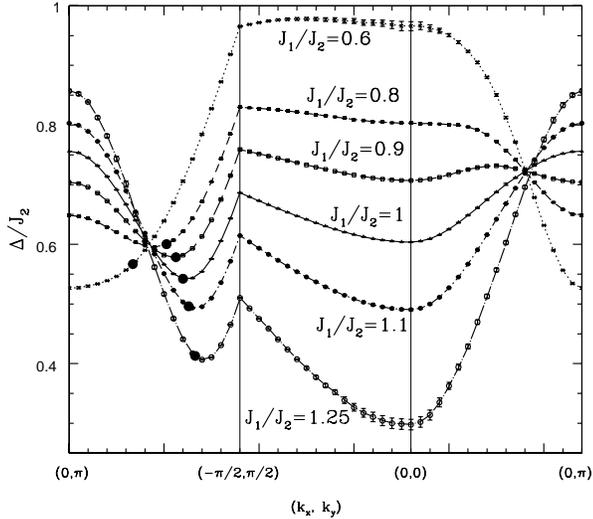,width=9cm}}}
%\par
\vspace{-1cm}
\caption{Plot of triplet excitation spectrum 
$\Delta(k_x, k_y)$ 
for the columnar dimerized $J_1-J_2$ model at $x=0.5$  
for coupling ratios
$J_1/J_2=0.6, 0.8, 0.9, 1, 1.1, 1.25$. The big circles 
indicate the locations of classical ordering wavevectors.
}
\label{fig_cdimer_mkp5}
\end{figure}
%=======================================================================

\subsection{Triplet excitation spectrum}

Fig. \ref{fig_Ising_mk} shows the triplet excitation spectra $\Delta(k_x, k_y)$
for $J_1/J_2=0,0.3,0.6$ obtained from the Ising expansions about Ne\'el order,
where we can see the energy gap vanishes at $(\pi,\pi)$ and its equivalent point.

We now present the spectra calculated from the columnar dimer expansions.
We have previously noted that the leading order columnar dimer expansions give
the minimum of the triplet excitation spectrum 
close to the classical ordering wavevector.
Fig. \ref{fig_cdimer_mkp5} shows the triplet excitation spectra 
$\Delta(k_x, k_y)$ 
for the columnar dimerized $J_1-J_2$ model at $x=0.5$  
and coupling ratios
$J_1/J_2=0.6, 0.8, 0.9, 1, 1.1, 1.25$. It can be seen
that the minimum gap  is located 
at $(0,\pi)$ for $J_1/J_2\lesssim 0.7$, or $(0,0)$
for $J_1/J_2\gtrsim 1.25$. For
 $J_1/J_2$ over the region $0.8-1.1$, the minimum
is close to the classical ordering wavevector.

%=======================================================================
\begin{figure}[h] %h: here; t:top of page; b:bottom of page; p: page of float
\vspace{-1cm}
%\par
\centerline{\hbox{\psfig{figure=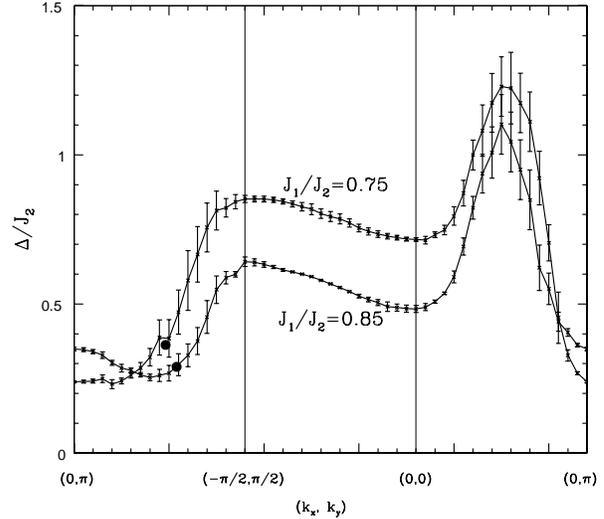,width=9cm}}}
%\par
\vspace{-1cm}
\caption{Plot of triplet excitation  spectrum $\Delta(k_x, k_y)$ 
derived from the columnar dimer expansions for the coupling ratios
$J_1/J_2=0.75, 0.85$ and $x=1$. The big full circles indicate the locations
of classical ordering wavevector.
}
\label{fig_cdimer_mk}
\end{figure}
%=======================================================================

\begin{figure}[t] %h: here; t:top of page; b:bottom of page; p: page of float
\vspace{-1.5cm}
%\par
\centerline{\hbox{\psfig{figure=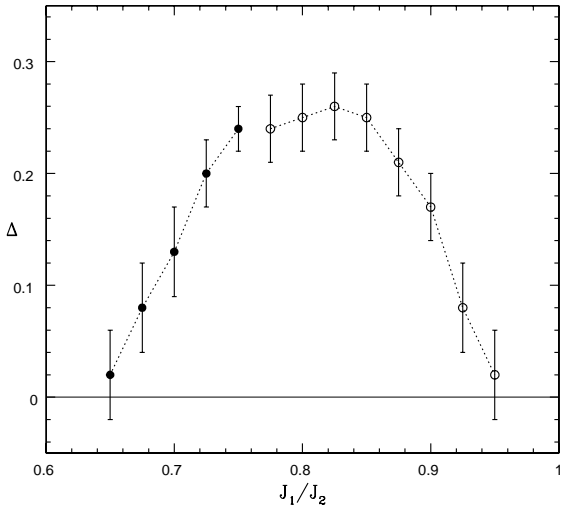,width=9cm}}}
%\par
\vspace{-1cm}
\caption{The minimum triplet excitation  gap $\Delta$ vs $J_1/J_2$
derived from the columnar dimer expansions. The full points indicate
the minimum gap is located at $(0,\pi)$.
}
\label{fig_cdimer_gap}
\end{figure}

Extrapolating the columnar dimer series 
to $x=1$ by the integrated differential approximants,
one can get the triplet excitation spectra for the fairly narrow parameter range
$0.7 \lesssim J_1/J_2 \lesssim 0.9$,
 and the results for $J_1/J_2=0.75$ and 0.85,
are shown in Fig. \ref{fig_cdimer_mk}. We find that the 
minimum gap is nonzero for these couplings. The minimum triplet
gap as a function of $J_1/J_2$ are given in Fig. \ref{fig_cdimer_gap}.
We find that up to $J_1/J_2\simeq 0.75$,
the minimum gap is located at $(0,\pi)$, and for $J_1/J_2 \gtrsim 0.75$,
the system is in an incommensurate phase, with the location of 
minimum gap given in Fig. \ref{fig_q}.

The excitation spectra for the regime of small $J_2/J_1$ will be
discussed in a later section.

%=======================================================================
%=======================================================================

\subsection{Phase Transitions and Critical Exponents}

In this section we describe various estimations of the
phase boundaries and the calculations of critical exponents
associated with the transitions. We note that the estimates
of critical exponents are not particularly accurate. The
uncertainties shown are a measure of the spread among the
Pad\'e approximants. Given the short length of the series, they are
best regarded as effective exponents.

First, combining the results of the Ising and the columnar dimer
expansions, we estimate that Ne\'el 
order disappears at $J_1/J_2=0.68(3)$. The phase adjacent to
that appears to be spontaneously dimerized in the columnar pattern.
This phase is analogous to the middle spontaneously 
dimerized phase in $J_{nn}-J_{nnn}$ square lattice model.\cite{zheng}
Our results are consistent with the transition from the Ne\'el
to the dimerized phase being a second-order transition, since, near
the critical point, the 
ground state energy from the columnar dimer expansion matches very
smoothly those from the Ising expansions.
For a first-order transition, there should be a
discontinuity in the slope at the transition.
Similarly, we estimate the transition point between spontaneously 
dimerized phase and the spiral phase to be at
$J_1/J_2 = 0.91(3)$.

\begin{figure}[h] %h: here; t:top of page; b:bottom of page; p: page of float
\vspace{-1cm}
%\par
\centerline{\hbox{\psfig{figure=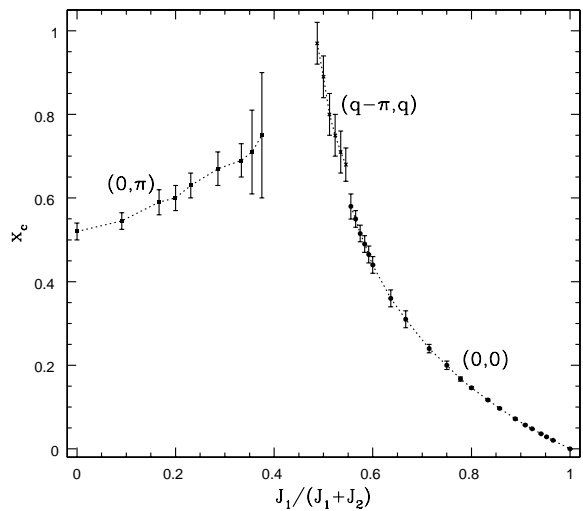,width=9cm}}}
%\par
\vspace{-1cm}
\caption{The critical point $x_c$ as a function of $J_1/(J_1+J_2)$
as obtained from Dlog Pad\'e approximants to the minimum energy 
gap of columnar dimer expansions. 
 The locations of minimum gap are $(0,\pi)$ for $J_1/J_2 \protect\alt 0.6$,
 or near classical  ordering  wavevector $(q-\pi,q)$ for
$0.95 \protect\alt J_1/J_2 \protect\alt 1.2$, or
(0,0) for $J_1/J_2\protect\gtrsim 1.25$.
}
\label{fig_cdimer_xc}
\end{figure}

Phase boundaries for the columnar dimerized Hamiltonian can 
be more fully studied by the columnar dimer expansions.
By applying the Dlog Pad\'e approximants to the triplet series
at the wavevector of the minimum gap,
we can locate the critical points at which the gap closes.
The results are presented in Fig. \ref{fig_cdimer_xc},
where in our analysis we have assumed that the minimum gap is located at
the classical ordering wavevector for
$J_1/J_2$ in the range $0.8-1.2$. 
The associated critical exponent $\nu$ is about $0.8(1)$, 
with not much variation with $J_1/J_2$.
It would be interesting to calculate the spectral weight for
the triplet excitations to see if they vanish, suggesting the
appearance of spinon excitations.\cite{spectral}
These results
are not accurate enough to decide if the transitions belong to the
O(3) nonlinear sigma model universality class ($\nu = 0.70$)\cite{leguillou}
or the O(4) exponents associated with the
SO(3) nonlinear sigma model ($\nu = 0.74$)\cite{chubukov,azaria}. 
For $J_1/J_2$ in the region $0.7-0.9$, we find critical points $x_c$
larger than unity. This is evidence that this phase is
spontaneously dimerized.

%=======================================================================
%=======================================================================

\begin{figure}[h] %h: here; t:top of page; b:bottom of page; p: page of float
\vspace{-1cm}
%\par
\centerline{\hbox{\psfig{figure=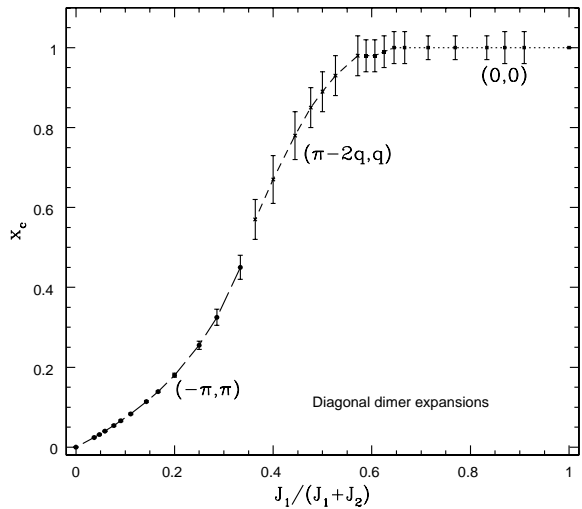,width=9cm}}}
%\par
\vspace{-1cm}
\caption{The critical point $x_c$ as a function of $J_1/(J_1+J_2)$ 
as obtained from Dlog Pad\'e approximants to the minimum energy gap of diagonal 
dimer expansions. 
The locations of minimum gap are $(0,0)$ for $J_2/J_1\protect\lesssim 0.75$,
or near classical  ordering  wavevector $(\pi-2q,q)$ for
$0.75 \protect\lesssim J_2/J_1 \protect\lesssim 1.75$, or
$(-\pi,\pi)$ for $J_2/J_1\protect\gtrsim 1.75$.
}
\label{fig_ddimer_xc}
\end{figure}

For diagonal dimer expansions, we can also get the phase 
boundary by using the Dlog Pad\'e approximants to 
locate the critical points where the gap vanishes.
The results are shown  in 
Fig. \ref{fig_ddimer_xc}. 
The associated critical exponent $\nu$ is 
found to be approximately 0.75(10), 
and also does not vary much with $J_2/J_1$.
For $J_2/J_1\protect\gtrsim 1.75$
the minimum of the energy gap is clearly $(-\pi, \pi)$, 
which is  the classical ordering
wavevector.

%=======================================================================
%=======================================================================

%=======================================================================
\begin{figure}[b] %h: here; t:top of page; b:bottom of page; p: page of float
\vspace{-1cm}
%\par
\centerline{\hbox{\psfig{figure=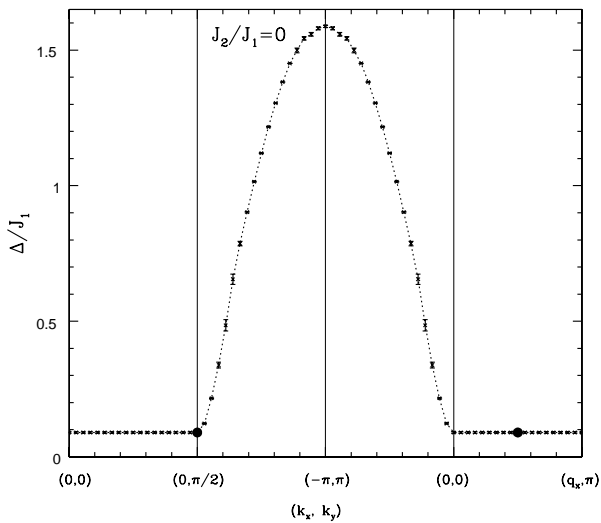,width=9cm}}}
%\par
\vspace{-1cm}
\caption{Plot of triplet excitation  spectrum 
$\Delta(k_x, k_y)$  along selected contours in Brillouin zone 
(derived from the diagonal dimer expansions) for coupling ratios
$J_2/J_1=0$. The big full circle indicates the position of the classical
ordering wavevector and $q_x=\pi (\pi-2q)/q=0$ for $J_2/J_1=0$.
}
\label{fig_ddimer_mk_p0}
\end{figure}
%=======================================================================

\subsection{Weakly Coupled Zig-Zag Chains}

In this section we discuss the excitation spectra for small $J_2/J_1$,
when the system can be thought of as weakly coupled spin-chains. The
coupling between the chains is frustrated, and has been called
zig-zag coupling in the literature.

%=======================================================================
\begin{figure}[h] %h: here; t:top of page; b:bottom of page; p: page of float
\vspace{-1.3cm}
%\par
\centerline{\hbox{\psfig{figure=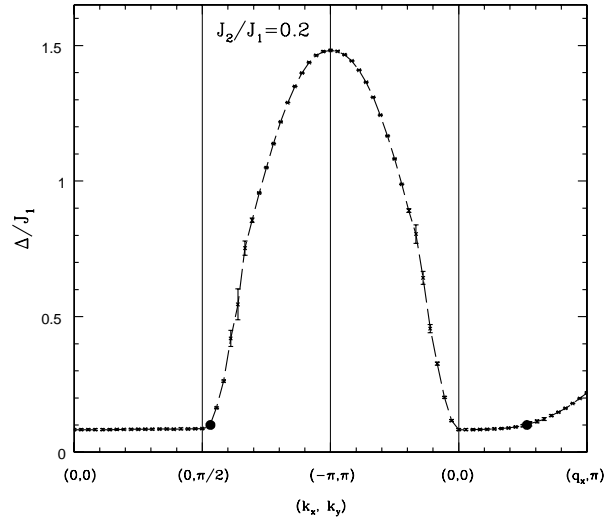,width=9cm}}}
%\par
\vspace{-1.5cm}
\caption{Plot of triplet excitation  spectrum 
$\Delta(k_x, k_y)$  along selected contours in Brillouin zone 
(derived from the diagonal dimer expansions) for coupling ratios
$J_2/J_1=0.2$. The big full circle indicates the position of the classical
ordering wavevector and $q_x=\pi (\pi-2q)/q=-0.12 \pi$ for $J_2/J_1=0.2$.
}
\label{fig_ddimer_mk_p2}
\end{figure}
%=======================================================================

\begin{figure}[h] %h: here; t:top of page; b:bottom of page; p: page of float
\vspace{-1.3cm}
%\par
\centerline{\hbox{\psfig{figure=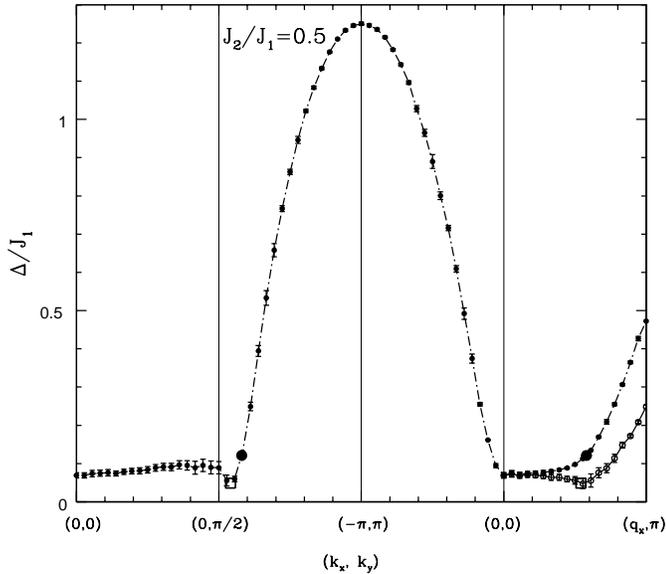,width=10cm}}}
%\par
\vspace{-1.5cm}
\caption{Plot of triplet excitation  spectrum 
$\Delta(k_x, k_y)$ along selected contours in Brillouin zone 
(derived from the diagonal dimer expansions) for coupling ratios
$J_2/J_1=0.5$. The big full point indicates the position of classical
ordering wavevector, while the big open square indicates the position of minimum gap.
The solid (open) points in rightmost windows are spectra along contour 
connecting $(0,0)$ and the classical ordering wavevector (the position
of minimum gap) by a straight line up to $k_y=\pi$.
}
\label{fig_ddimer_mk_p5}
\end{figure}

For small values of $J_2/J_1$, 
the critical points in the diagonal dimer expansions
lie very close to unity, and 
it is very difficult
to locate the wavevector of the gap minimum at $x=1$. 
Fig. \ref{fig_ddimer_mk_p0} gives the dispersion at $x=1$ for the case 
$J_2=0$ (i.e., decoupled  spin chains).
Here we get a small but nonzero minimum gap, due to the fact that
we have not taken into account the singular behavior
of the spin-chain. This can be
done by extrapolating the series in a new variable
\begin{equation}
\delta = 1 - (1 - x)^{2/3},
\end{equation}
in which case the gap does vanish. For $J_2$ non-zero, we
do not know if $x=1$ is gapped or critical, and hence do not
know which extrapolation variable to use. For simplicity,
we discuss primarily an extrapolation in the original variable,
which assumes there is no singularity at $x=1$.
Note, however, that even with this extrapolation
we do get a flat dispersion between $(0,0)$ and $(0,\pi/2)$, as expected. 
As we increase $J_2/J_1$ (up to $J_2/J_1\simeq 0.2$), we still get a 
very flat dispersion over the region 
$(0,0)$ and $(0,\pi/2)$. The gap over this region is also 
very small (similar to that for $J_2=0$),
and in fact it is zero within errorbars if we extrapolate the series using
the variable $\delta$. As an example,
the dispersion is shown in Fig. \ref{fig_ddimer_mk_p2}
for $J_2/J_1=0.2$
(extrapolated assuming no critical points for $x<1$).
At larger values of $J_2/J_1$, for example for $J_2/J_1=0.5$,
the dispersion is more pronounced and is shown in Fig. \ref{fig_ddimer_mk_p5}.
We can see that the position of the minimum gap is not located exactly
at the classical ordering wavevector, but deviates from it
along the direction connecting $(0,\pi/2)$ and $(-\pi,\pi)$.
This location of the minimum gap 
is plotted in Fig. \ref{fig_q}.

%=======================================================================
%=======================================================================

The wavevector $q$ obtained from both columnar dimer expansions and diagonal dimer
expansions can be fitted very well by the simple form:
\begin{equation}
q = \cases {\arccos [-(3 J_2/J_1 -2)/2], & $ 0.75 \gtrsim J_1/J_2\le 1$;\cr
\arccos \{ -[2(3 J_1/J_2 -2)]^{-1} \}, & $J_1/J_2 \ge 1 $.\cr} \label{eq_qfit}
\end{equation}
which is shown by a dashed lined in Fig. \ref{fig_q}.

Using the above results for $q$, we can obtain more reliable results for the
ground state energy by the Ising expansion, and the results 
are shown in Fig. \ref{fig_e0}.
The results for the magnetisation $M$ obtained
using this wavevector agree
(within errorbars)  
with the results obtained using the classical wavevector,
shown in Fig. \ref{fig_M} and, hence, are not shown.

\section{Conclusions}
In this paper we have studied a class of Heisenberg models, which
interpolate between the square-lattice, the triangular-lattice and
the linear-chain models. 
As discussed in the introduction, these models should
describe the magnetic properties of
the insulating phase of a range of layered
organic superconductors. Furthermore, 
using uniaxial stress it may be possible to vary
the parameter $J_1/J_2$ and induce some of the 
quantum phase transitions considered in this paper.
We find that the model has a particulary rich phase diagram:
Ne\'el, spiral,
dimerized, and possible critical phases
occur. Except, perhaps, for the region of weakly coupled
chains the phase diagram is rather well determined by our
series expansions, and it would be difficult for any other
numerical method to have comparable accuracy. 
Some of the most notable features are that the two-sublattice Ne\'el
phase appears stable beyond its classical regime of stability and
in the spiral phase, the ordering wavevector is different from
the classical ordering wavevector except for the case of
the triangular-lattice Heisenberg model. The region of weakly coupled
chains is nearly critical. In studying the
instability of the dimerized phase to the spiral phase, the critical exponents
for various dimer expansions give values of $\nu$ in the range $0.65<\nu<0.9$
which is consistent with either $O(4)$ or $O(3)$ universality class. This,
as well as the possibility of spinon excitations in the 
quantum disordered phases of this model, deserve
further attention.

\acknowledgments
This work is supported in part by a grant
from the National Science Foundation (DMR-9616574) (R.R.P.S), the
Gordon Godfrey Bequest for Theoretical Physics at the University of
NSW, and by the Australian Research Council (Z.W. and R.M.).
The computation has been performed on Silicon Graphics Power 
Challenge and Convex machines. We thank the New South Wales 
Centre for Parallel Computing for facilities and assistance
with the calculations. 
We thank J. Merino, J. Oitmaa, C.J. Hamer and S. Sachdev 
for helpful discussions. We thank C.J. Hamer and J. Oitmaa
for providing some of the computer codes we used.

% \newpage

\end{document}